\documentclass[a4paper,11pt]{article}
\pdfoutput=1 

\usepackage{jcappub} 
\usepackage[T1]{fontenc} 
\usepackage{graphicx}
\usepackage{xcolor}
\usepackage{tikz}
\usepackage{overpic}
\usepackage{rotating}
\usetikzlibrary{arrows,shapes,chains}

\def\be{\begin{equation}}
\def\ee{\end{equation}}

\def\ba{\begin{align}}
\def\ea{\end{align}}

\def\bc{\begin{cases}}
\def\ec{\end{cases}}

\title{Enhance Primordial Black Hole Abundance through the Non-linear Processes around Bounce Point}
\author[a, b]{Jie-Wen Chen,}
\author[c, d]{Mian Zhu,}
\author[e, f]{Sheng-Feng Yan,}
\author[g, h]{Qing-Qing Wang,}
\author[g, h]{Yi-Fu Cai}

\affiliation[a]{MOE Key Laboratory of Fundamental Physical Quantities Measurements, School of Physics,
Huazhong University of Science and Technology, Wuhan, Hubei 430074, China}
\affiliation[b]{Department of Astronomy, School of Physics, Huazhong University of Science and Technology, Wuhan, Hubei 430074, China}
\affiliation[c]{Department of Physics, The Hong Kong University of Science and Technology, Clear Water Bay, Hong Kong S.A.R., China}
\affiliation[d]{HKUST Jockey Club Institute for Advanced Study, The Hong Kong University of Science and Technology, Clear Water Bay, Hong Kong S.A.R., China}
\affiliation[e]{Istituto Nazionale di Fisica Nucleare (INFN), Sezione di Milano, Via Celoria 16, 20146, Milano, Italy}
\affiliation[f]{DiSAT, Università degli Studi dell'Insubria, Via Valleggio 11, 22100, Como, Italy}
\affiliation[g]{CAS Key Laboratory for Researches in Galaxies and Cosmology, Department of Astronomy, School of Physical Sciences, University of Science and Technology of China, Hefei, Anhui 230026, China}
\affiliation[h]{School of Astronomy and Space Science, University of Science and Technology of China, Hefei, Anhui 230026, China}

\emailAdd{chjw@hust.edu.cn, mzhuan@connect.ust.hk, Shengfeng.Yan@mi.infn.it, wangqqq@mail.ustc.edu.cn, yifucai@ustc.edu.cn}

\abstract{The non-singular bouncing cosmology is an alternative paradigm to inflation,
wherein the background energy density vanishes at the bounce point, in the context of Einstein gravity. 
Therefore, the non-linear effects in the evolution of density fluctuations ($\delta \rho$) may be strong in the bounce phase,
which potentially provides a mechanism to enhance the abundance of primordial black holes (PBHs). 
This article presents a comprehensive illustration for PBH enhancement due to the bounce phase. 
To calculate the non-linear evolution of $\delta \rho$, the Raychaudhuri equation is numerically solved here. 
Since the non-linear processes may lead to a non-Gaussian probability distribution function for $\delta \rho$ after the bounce point, 
the PBH abundance is calculated in a modified Press-Schechter formalism. 
In this case, the criterion of PBH formation is complicated, 
due to complicated non-linear evolutionary behavior of $\delta \rho$ during the bounce phase. 
Our results indicate that the bounce phase indeed has potential to enhance the PBH abundance sufficiently. 
Furthermore, the PBH abundance is applied to constrain the parameters of bounce phase,
providing a complementary to the surveys of cosmic microwave background and large scale structure. 
}

\begin{document}
\maketitle
\flushbottom

\section{Introduction}\label{Sec:Intro}

Primordial black holes (PBHs) are believed to originate from extremely over-dense regions in the early Universe \citep{Zel'dovich&Novikov1967, Hawkin1971, Carr&Hawking1974}.
Unlike astrophysical black holes, which evolve from massive stars and contain masses $\gtrsim 5 ~{\text M}_\odot$ \citep{Ozel_2010},
PBHs can be formed much earlier than the birth of first-generation stars, 
and their masses are allowed to distribute in a very wide range ---
in principal, from Planck mass ($\sim 10^{-5}$ g) to the mass of observable Universe ($\sim 10^{55}$ g) \citep{Carr:2010, Sasaki:2018, Carr:2020}.
In view of this, there are various motivations to introduce PBHs in cosmology and astrophysics.
For example, the massive PBHs ($>10^{15}$ g) can serve as
a candidate for cold dark matter \citep{IvanovPRD94, Carr:2016drx}
and seeds of super massive black holes at high redshifts \citep{Carr:1984, Bean:2002}. 
Moreover, the light PBHs ($<10^{15}$ g) have strong Hawking radiations \citep{Page&Hawking:1976, Carr:1976},
so they may be responsible for some electromagnetic emission phenomena \citep{Carr:2010, Belyanin:1996, Miyama:1978, Ackermann_2018}.

The attentions on PBHs keep stupendously increasing in recent years,
especially since LIGO and Virgo collaborations achieve fruitful successes on gravitational wave (GW) detection.
LIGO and Virgo so far have announced tens of GW events, 
and have also brought several potential evidences for PBHs \citep{Bird:2016, Hall:2020, Luca:2020, Franciolini2021, LIGO-PBH:2021}.
In particular, for the event GW190521 \citep{LIGO-Virgo-GW190512}, 
one black hole's mass $85^{+21}_{-14} ~{\text M}_\odot$ is believed to reside in the $(60-130)~{\text M}_\odot$ mass gap,
which is forbidden in stellar evolution theories \citep{Woosley2017}. 
The literature \cite{DeLuca:2020sae} shows that this black hole may have a primordial origin,
if the PBHs can accrete efficiently before reionization epoch. 
In addition, although the events GW200105 and GW200115 are officially announced as GW signals from neutron star-black hole binaries \citep{Abbott_2021},
they are also compatible with the scenario of PBH binaries \citep{Wang:2021iwp}. 
Moreover, other GW detection programs, including NANOGrav \citep{NANOGrav-PBHs:2021}, LISA \citep{LISA:2017}, Taiji \citep{Taiji:2017}, TianQin \citep{Mei:2020}, Gaia and THEIA \citep{Gaia:2021} etc., have all treated PBHs as a potential target. 
Therefore, one can expect that the requirement for PBH researches will become more and more emergent in the epoch of GW astronomy.

The abundance and mass function of PBHs rely on power spectrum and probability distribution function (PDF) of primordial density fluctuations \citep{Carr:1975}.
In the literature, a Gaussian PDF is generally assumed, which is named Press-Schechter formalism \citep{Press-Schechter:1974}. 
In this case, a power spectrum with extra enhancement on certain length scales (much smaller than the scales of cosmic microwave background (CMB) surveys $1-10^4~{\rm Mpc}$ \citep{Planck:2018vyg}) 
is required to generate PBHs with certain masses \citep{Bellido-Linde-Wands:1996, Bellido:2017,  Carr:2017, Cai:2018, Chao_2019, Chao:2020, Zhou:2020}.
To achieve the enhancement, particular mechanisms in the early Universe are introduced, such as hybrid inflation \citep{Bellido-Linde-Wands:1996}, inflection-point inflation \citep{Bellido:2017}, and sound speed resonance of the fluctuations \citep{Cai:2018, Chao_2019, Chao:2020, Zhou:2020} etc..
Moreover, recent literature \cite{Young_2013, Franciolini_2018, Passaglia2019, Luca_2019, Atal:2019, Yoo_2019, CaiPiSasaki:2019, Atal_2020, Braglia_2020, Kitajima_2021, Riccardi_2021, Cai:2021zsp,Cai:2022erk} has pointed out that the extremely compact regions
should inevitably have highly non-linear evolution, i.e. obeying a non-Gaussian PDF.   
Therefore, the non-Gaussianity (NG) can have a significant impact on the PBH abundance. 

Since inflation is currently the most widely accepted paradigm of early Universe,
most PBH researches are based on inflationary scenario.
Meanwhile, the bouncing cosmology \citep{Wands:1999, Finelli:2002, Brandenberger:2017, BATTEFELD20151} 
can satisfy the CMB constraints \citep{Planck:2018vyg} as well,
hence it is deemed as an alternative scenario to inflation. 
In the non-singular bouncing scenario, the Universe typically starts with a contracting phase,
then turns to expand when the scale factor is small enough (but still larger than $0$), 
and finally evolves to the expanding Big Bang phase.  
Usually, the moment of the minimal scale factor is called bounce point \citep{Brandenberger:2017}.
So far, only few works \citep{Quintin_2016, CHEN2017561} have investigated the PBH formation in bouncing cosmology  \citep[see][as well]{CARR&COLEY2011},
and the analyses therein concentrate on the contracting phase, not including the bounce point.
The results of PBH abundance given by these researches seem pessimistic.
For example, our previous work \cite{CHEN2017561} shows that the density fluctuation given by matter-contracting phase does not enhance the PBH abundance significantly, 
unless the phase is physically disfavored, e.g. the Hubble parameter is close to or larger than Planck scale.

However, it is inspiring to note that,
the background density ($\bar \rho$) around the bounce point approaches $0$,
in context of Einstein gravity \citep{CHEN2017561, Cai:2012va, Quintin:2015rta}. 
This implies that the density  fluctuation ($\delta \rho/\bar \rho$) may diverge \footnote{It is an unphysical divergent as long as $\delta \rho$ is limited. The realization of bounce phase without pathology can be found in the literature  \cite{Cai:2016thi,Cai:2017dyi,Mironov:2018oec,Kolevatov:2018mhu,Mironov:2019mye,Volkova:2019jlj,Ilyas:2020qja,Mironov:2022ffa} }, 
and the evolution of $\delta \rho$ can be highly non-linear and the PDF may have strong NGs. 
These non-linear effects/NGs have been confirmed at least for some specific cases \citep{Quintin:2015rta}. 
Therefore, the non-linear processes around bounce point naturally provide a mechanism to increase PBH abundance,
and we will illustrate it in this article.

The rest of this article is organized as follows.
In \autoref{Sec:Bounce}, we briefly review the non-singular bouncing scenario and present a model-agnostic parametrization of the bounce phase.
In \autoref{Sec:Fluctuations}, we introduce the Rychaudhuri equation to numerically calculate the non-linear evolution of density fluctuation in the bounce phase.
In \autoref{Sec:PBHs}, we evaluate the PBH abundance in the bouncing scenario, and then 
use the PBH abundance to constrain the parameters of bounce phase. 
Finally, we summarize in \autoref{Sec:Conclusion}.
Throughout this paper, the Planck unit with $c=\hbar=8\pi G=1$ is adopted, unless otherwise specified.
Under this convention, the Planck mass is $M_p=(8\pi G)^{-1/2}=4.6\times 10^{-6}$ g,
and the present Hubble parameter is $H_0=67.4{\rm~km~s^{-1}~Mpc^{-1}}=5.9\times 10^{-61}$ \citep{Planck:2018vyg}.

\section{A Brief Description of Non-singular Bounce Phase}\label{Sec:Bounce}

The non-singular bouncing scenario can be realized by multiple mechanisms \citep{Brandenberger:2017, BATTEFELD20151},
including those modifying the General Relativity (GR) \citep{Cai_2009-NGRbounce, Gao_2010, Cai_2011-f(T)bounce, Haro:2013, Cai_2014-loopbounce, Cai_2016-f(T)bounce} and not \citep{Cai_2007-qb, Cai-2009-LWbounce, RadBounce:2010, Lin_2011, Qiu_2011, Easson_2011, Fermi-bounce:2014, ALEXANDER201597, ekpyrobounces:2019}.
In this article, we only consider the context of GR 
or alternative gravity theories which are equivalent to GR with a dynamical fluid. 
In this case,
the background dynamics can be described by the Friedmann-Lema\^\i tre-Robertson-Walker (FLRW) equation
\be
\label{eq:FLRW}
\bc
\bar H^2=\frac{1}{3} \bar \rho \\
\dot {\bar H}+\bar H^2=-\frac{1}{6} (\bar \rho +3\bar P)
\ec. 
\ee
Here $\bar H=\dot {\bar a}/\bar a$, $\bar \rho$ and $\bar P$ denote background Hubble parameter, density and pressure, respectively. 
The $\bar a$ means the background scale factor.
The "$\bar~$" denotes a variable for the background fluid, and the "$\dot~$" means derivative with respect to background time $t$. 
Furthermore, the background pressure is usually parameterized as $\bar P=w \bar \rho$, with $w$ named equation-of-state (EOS) parameter.

\begin{figure}
  \centering
  \begin{overpic}[width=0.7\textwidth]{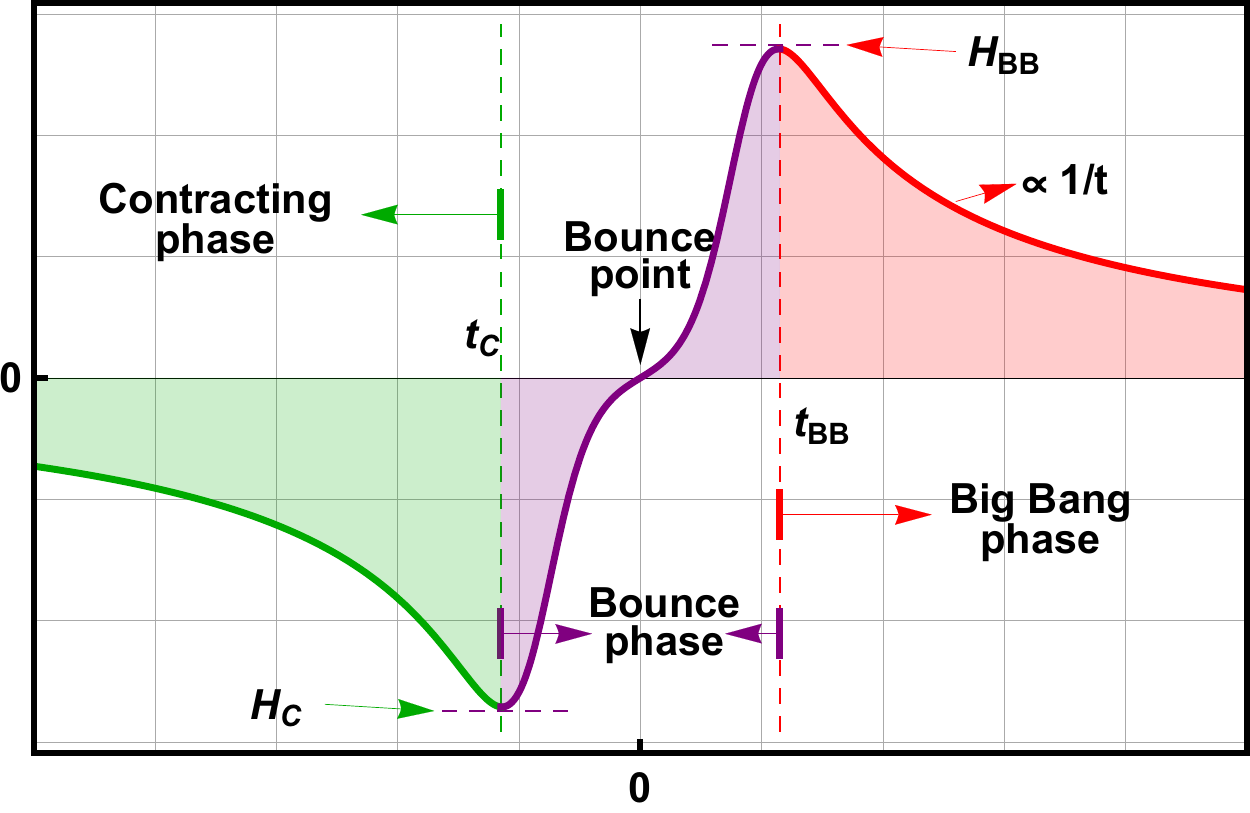}
  \put(50,-5){$t$} \put(-8,32){\begin{sideways} $\bar H(t)$ \end{sideways}}
  \end{overpic}
  ~\\
  ~\\
  \caption{An illustration for the evolution of $\bar H(t)$ throughout a bouncing cosmology.}\label{Fig:BounceFig}
\end{figure}

As mentioned in \autoref{Sec:Intro}, the Universe starts with a contracting phase,
during which $\bar H$ is negative and $w$ is generally larger than $-1/3$ \citep{Brandenberger:2017}. 
Subsequently, the bounce phase takes place when the Universe is small enough to avoid the cosmological singularity, 
hence the null energy condition (NEC) $\bar \rho+\bar P \geq 0$ should be violated, according to the singularity theorems \citep{Hawking-Penros-Bondi}. 
This indicates that the constraints $w<-1$ and $\dot {\bar H}>0$ are required during the bounce phase. 
Additionally, the condition $\bar H=0$ holds at the bounce point \citep{Brandenberger:2017}. 
Given above, the evolutionary behavior of $\bar H$ during the bounce phase can be sketched as follows ---
$\bar H$ is negative before bounce point, vanishes at the point, and becomes positive after it (see \autoref{Fig:BounceFig} as well). 
The final stage of the Universe is the standard Big Bang phase,
during which the background dynamics can be simply described by the spatially flat $\Lambda$CMD model 
\be
\label{eq:LCDM}
\bar H(z)=H_0 \sqrt{\Omega_\Lambda+\Omega_m(1+z)^3+\Omega_m(1+z_{\rm eq})^{-1}(1+z)^4},
\ee
where $z\equiv 1/\bar a-1$ is the cosmological redshift, 
$H_0\simeq 67.4~ {\rm km s}^{-1}{\rm Mpc}^{-1}$ denotes the present Hubble parameter,
$\Omega_m \simeq 0.315$ stands for density fraction for matter component today,
$\Omega_\Lambda \simeq 1-\Omega_m$ represents density fraction for dark energy today,
and $z_{\rm eq}\simeq 3400$ is the redshift when matter and radiation have equal densities \citep{Planck:2018vyg}.

Since the PBH formation in contracting phase have been studied in \cite{Quintin_2016, CHEN2017561}, 
we concentrate our consideration on the bounce phase in this article, 
i.e. from the end of contracting phase ($t_{\rm C}$) to the beginning of Big Bang phase ($t_{\rm BB}$). 
Hereafter, the Hubble parameters at $t_{\rm C}$ and $t_{\rm BB}$ are taken as two free parameters $H_{\rm C}$ and $H_{\rm BB}$.
For not losing generality,
we adopt a model-agnostic parametrization of $\bar H$ during bounce phase, 
i.e. Taylor expanding it as 
\be
\label{eq:bounce}
\bar H(t)= {\mathop \sum \limits_{\gamma=1}^{\infty}} \Upsilon_\gamma t^\gamma, ~~~~~~~t_{\rm C}\leq t \leq t_{\rm BB}.
\ee
where the coefficients $\Upsilon_\gamma$ are parameters to be determined, 
and $t=0$ is set as the bounce point (see \autoref{Fig:BounceFig}). 
It is seen that to ensure the condition $\dot {\bar H}>0$ during the bounce phase, 
$\bar H$ should be dominated by one or several terms of odd $\gamma$. 
For simplicity, 
we only consider one odd term in this work, and \autoref{eq:bounce} reduces to 
\be
\label{eq:bounce-H}
\bar H(t) \simeq \Upsilon_\gamma t^\gamma, ~~~~~~~t_{\rm C}\leq t \leq t_{\rm BB}, 
\ee
with $\gamma$ being an odd number (e.g. 1, 3, 5...).
It is important to note that, although the parametrization of \autoref{eq:bounce-H} is not as generic as \autoref{eq:bounce}, 
it still applies for a variety of bouncing models \citep{Cai_2007-qb, Cai:2012va, Quintin:2015rta, Quintin:2014prd, MEMBIELA2014196, Sriramkumar_2015}.
As a result, the e-folding number with respect to the bounce point can be calculated as
\be
\label{eq:bounce-n}
n \equiv \int_0^t \bar H(t') dt'=\frac{\Upsilon_\gamma t^{\gamma+1}}{\gamma+1}=\frac{|H|^{\frac{\gamma+1}{\gamma}}}{(\gamma+1)\Upsilon_\gamma^{1/\gamma}}, 
\ee
and the scale factor is 
\be
\label{eq:bounce-a}
\bar a (t) \equiv a_{\rm b} e^{n}=a_{\rm b} \exp \left(\frac{|H|^{\frac{\gamma+1}{\gamma}}}{(\gamma+1)\Upsilon_\gamma^{1/\gamma}}\right), 
\ee
where $a_b$ denotes the scale factor at the bounce point.

At the end of this part, we give a short guidance on how to practically calculate the background dynamics during the bounce phase, using our parametrization in \autoref{eq:bounce-H}. 
In our treatment, we set $H_{\rm C}$, $H_{\rm BB}$, $\gamma$, $n_{\rm BB}$ as four independent parameters, 
with the last one denoting the e-folding number at $t_{\rm BB}$. 
Once the four parameters are given, the evolution of $\bar a(t)$ and $\bar H(t)$ will be determined as follows.  
Firstly, taking these parameters into \autoref{eq:bounce-H} and \autoref{eq:bounce-n}, 
one obtains the values of $t_{\rm C}$, $t_{\rm BB}$ and $\Upsilon_m$. 
Furthermore, taking $H_{\rm BB}$ into Eq. \autoref{eq:LCDM} and numerically solve the equation,
one can get the scale factor at the beginning of the Big Bang phase (denoted as $\bar a_{\rm BB}$, hereafter).
Hence, the scale factor at the bounce point is known as $a_b=a_{\rm BB} \exp(-n_{\rm BB})$.
Finally, $\bar a$ and $\bar H$ can be calculated from \autoref{eq:bounce-a} and \autoref{eq:bounce-H}, respectively.
We notice that the our parametrization may lead to discontinuities of $\dot {\bar H}$ and $w$ at the joint points $t_{\rm C}$ and $t_{\rm BB}$, but the discontinuities can be allowed in this work. 
Furthermore, the values of the above parameters may be constrained in some specific bouncing models, we do not consider those constraints here.

\section{Density Fluctuations During Bounce Phase}\label{Sec:Fluctuations}
\subsection{Raychaudhuri Equation}\label{Sec:Raychaudhuri-Eq}

To investigate the non-linear evolution of density fluctuations during the bounce phase,
we introduce the Raychaudhuri equation \citep{Raychaudhuri, Raychaudhuri:1979, liddle_lyth_2000},
which describes fluctuations with isotropic stress in a comoving gauge
\be
\label{eq:Raychaudhuri}
\frac{dH}{dt_c}+H^2=-\frac{1}{6}(\rho+3P)-\frac{1}{3}\nabla \cdot \frac{\nabla P}{\rho+P},
\ee
where $H$, $\rho$ and $P$ denote the fluctuated Hubble parameter, energy density and pressure, respectively.
The $t_c$ is the fluctuated time (proper time along a comoving worldline \citep{liddle_lyth_2000}). 
Typically, the density and pressure are expanded as
\be
\bc
\rho=\bar \rho+\delta \rho=\bar \rho(1+\delta)\\
P=\bar P+\delta P=\bar \rho (w+c_s^2 \delta)
\ec,
\ee
with $\delta \rho$, $\delta$, $\delta P$ and $c_s$ being density fluctuation, density contrast, pressure fluctuation and adiabatic sound speed, respectively.
The fluctuated time $t_c$ can be achieved by  \citep{liddle_lyth_2000}
\be 
\label{eq:comoving}
\nabla^2 \left(\frac{dt}{dt_c} \right)=\nabla^2 \left(\frac{\delta P}{\rho+P} \right).
\ee
Additionally, to calculate the evolution of fluctuations, one also requires a continuity equation \citep{liddle_lyth_2000}
\be
\label{eq:continuity}
\frac{d \rho}{d t_c}= -3 H(\rho+P).
\ee

In the following, we will calculate the evolution of the fluctuated regions around bounce point using  \autoref{eq:Raychaudhuri}-\autoref{eq:continuity}. 
Let us start with making some simplifications.  
Firstly, all the fluctuated regions are considered as isolated and spherical clouds or voids
embedded in the background fluid. 
This assumption has been widely adopted in the Press-Schechter formalism \citep{Press-Schechter:1974}. 
Hence, \autoref{eq:comoving} reduces to
\be
\label{eq:comoving-2}
\frac{dt}{dt_c}\equiv \xi =1+\frac{\delta P}{\rho+P}=1+\frac{c_s^2 \delta \rho}{\rho+P}.
\ee
Secondly, the approximations $|\nabla \delta \rho| \simeq |\delta \rho/R|$ and $|\nabla^2 \delta \rho| \simeq |\delta \rho/R^2|$ are taken, with $R$ being the physical radius of the fluctuated region.
Thirdly, we set $c_s$ as a constant for both time and spatial coordinates, during the bounce phase.
Based on the above assumptions, the gradient operator can be simply replaced by $\nabla \rightarrow i k/a$
when it acts linearly on $\delta \rho$, 
where $a$ is the fluctuated scale factor defined by $d \ln a/dt_c \equiv H$,
and $k=a/R$ the comoving wavenumber of the fluctuated region. 
Moreover, according to \autoref{eq:continuity},
the scale factor can be expressed as 
\be
\label{eq:a}
a= a_{\rm C} \exp \left(-\frac{1}{3}\int \frac{d\rho}{\rho+P} \right),
\ee
with $a_{\rm C}$ being the background scale factor at the initial moment $t_{\rm C}$. 

As a result, \autoref{eq:Raychaudhuri} reduces to 
\be
\label{eq:Raychaudhuri-2}
 -\frac{\xi \dot \xi \dot \rho+\xi^2 \ddot \rho}{\rho+P}+\frac{4\xi^2 \dot \rho^2+3\xi^2\dot\rho \dot P}{3(\rho+P)^2}=-\frac{1}{2}(\rho+3P)+\frac{c_s^2 k^2}{a^2} \left[\frac{\delta \rho}{\rho+P}-\frac{(1+c_s^2)\delta \rho^2}{(\rho+P)^2} \right],
\ee
with $\xi$ given by \autoref{eq:comoving-2}, $a$ given by \autoref{eq:a}, and the variables $\bar \rho$ and $\bar P$ given by  \autoref{eq:FLRW}, respectively. 
By numerically solving \autoref{eq:Raychaudhuri-2}, the non-linear evolution of $\delta \rho$ can be calculated, 
which will be discussed in the following parts. 

\subsection{Linear Approximations}\label{Sec:linear}

Since the non-linear evolution of $\delta \rho$ is significant only around the bounce point, 
a linear dynamics remains valid during the rest epochs.  
Therefore, the linear approximation of \autoref{eq:Raychaudhuri-2} will be studied firstly, 
which can greatly simplify the analysis and provide initial conditions for the non-linear computations. 



Up to the linear order of $\delta \rho$, \autoref{eq:comoving-2} reduces to $\xi=1+c_s^2 \delta \rho/(\bar \rho+\bar P)$ and
the scale factor can be treated as unperturbed ($a=\bar a$). Therefore, \autoref{eq:Raychaudhuri-2} reduced to
\be
\label{eq:linear}
\ddot {\delta \rho}+\left( 5\bar H-\frac{\ddot {\bar H}}{\dot {\bar H}} \right)\dot {\delta \rho}+\left( \frac{c_s^2 k^2}{\bar a^2}+6\bar H^2+4 \dot {\bar H}-\frac{3\bar H \ddot{\bar H}}{\dot{\bar H}} \right)\delta \rho=0.
\ee
Inserting the background dynamics of the bounce phase \autoref{eq:bounce-H}, one obtains
\be
\label{eq:linear-m}
\ddot {\delta \rho}+\left(5\Upsilon_\gamma t^\gamma -\frac{\gamma-1}{t}\right) \dot {\delta \rho}+\left[ \frac{c_s^2 k^2}{\bar a^2}+(3+\gamma)\Upsilon_\gamma t^{\gamma-1}+6 \Upsilon_\gamma^2 t^{2\gamma}  \right]\delta \rho=0,
\ee
with $\gamma$ being an odd number.
It is seen that $\delta \rho$ has an oscillatory behavior during bounce phase, even for the long-wavelength modes $k \rightarrow 0$. 
For example, in the case $\gamma=1$ and $t \rightarrow 0$, \autoref{eq:linear-m} reduces to an equation of harmonic oscillator $\ddot {\delta \rho}+4 \Upsilon_1\delta \rho=0$. 
This means that an initial void may become a cloud during the bounce phase, and vise versa. 
Therefore, the probability of PBHs originated from both initial clouds and voids should be considered in the following numerical computations.

\begin{figure}
\centering
   \tikzstyle{box}=[rectangle, rounded corners, draw, thick, fill=blue!5 ]
\tikzstyle{box0}=[rectangle, rounded corners, draw, thin, fill=white]
\tikzstyle{box2}=[rectangle, rounded corners, draw, thick, minimum width=100, minimum height=60, fill=blue!5]
\tikzstyle{box3}=[rectangle, rounded corners, draw, dashed, thick, minimum width=300, minimum height=180, fill=green!5]
\tikzstyle{box4}=[rectangle, rounded corners, draw, thin, minimum width=100, minimum height=60, fill=white]
\tikzstyle{dia}=[diamond, draw, thin, aspect=3,  text centered, fill=white]
\tikzstyle{dia2}=[diamond, draw, thin, aspect=2,  text centered, fill=white]
\tikzstyle{arrow} = [thick,->,>=stealth, orange]
\begin{tikzpicture}
\node at (8,-1.5) [box3] {};
\node at (8,1.3) [] {\textbf{iteration:}};
\node (step1) at (0, 0) [box] {$a^{(0)}(t)=\bar a (t)$};
\node (step1b) at (2, 0.5) [ ] {\textcolor{blue}{$i=0$}};
\node (step2) at (4, 0) [box] {$a^{(i)}(t)$};
\node (step2b) at (6, 0.5) [ ] {\autoref{eq:Raychaudhuri-2}};
\node (step3) at (8, 0) [box] {$\delta \rho^{(i)}(t)$};
\node (step2b) at (10, 0.5) [ ] {\autoref{eq:a}};
\node (step4) at (12, 0) [box] {$a^{(i+1)}(t)$};
\node (step5) at (8,-3) [dia] {~~~~~~~~~~~~~~~~~~~~~~~~~~~~~};
\node (step5b) at (8,-3) []{$\left| \frac{a^{(i+1)}(t)}{a^{(i)}(t) }-1 \right| < \epsilon$};
\node (step6b) at (5, -1.2) [] {\textcolor{blue}{$i=i+1$}};
\node (step6) at (5.75, -1.6) [] {\textcolor{blue}{$a^{(i)}(t)=a^{(i+1)}(t)$}};
\node (result) at (0, -3) [box2] {~~~~~~};
\node (result0) at (0, -2.25) [] {\textbf{results:}};
\node (result1) at (0, -3) [] {$a(t)=a^{(i)}(t)$};
\node (result2) at (0, -3.5) [] {$\delta \rho(t)=\delta \rho^{(i)}(t)$};
\node  at (8.5,-4) [] {\textcolor{blue}{Yes}}; 
\draw[arrow](step1) -- (step2);
\draw[arrow](step2) -- (step3);
\draw[arrow](step3) -- (step4);
\draw[arrow](step4) --(12,-1.5)--(8,-1.5)-- (step5);
\draw[arrow](step5) --(4,-3)--(step2);
\draw[arrow](step5)--(8, -4.5)--(3,-4.5)--(3,-3)--(result);
\end{tikzpicture}
\caption{Process of the iteration in computing $a$ from \autoref{eq:a} and $\delta \rho$ from \autoref{eq:Raychaudhuri-2}. Here the precision parameter $\epsilon$ is taken as $0.1$.}
    \label{fig:flow}
\end{figure}
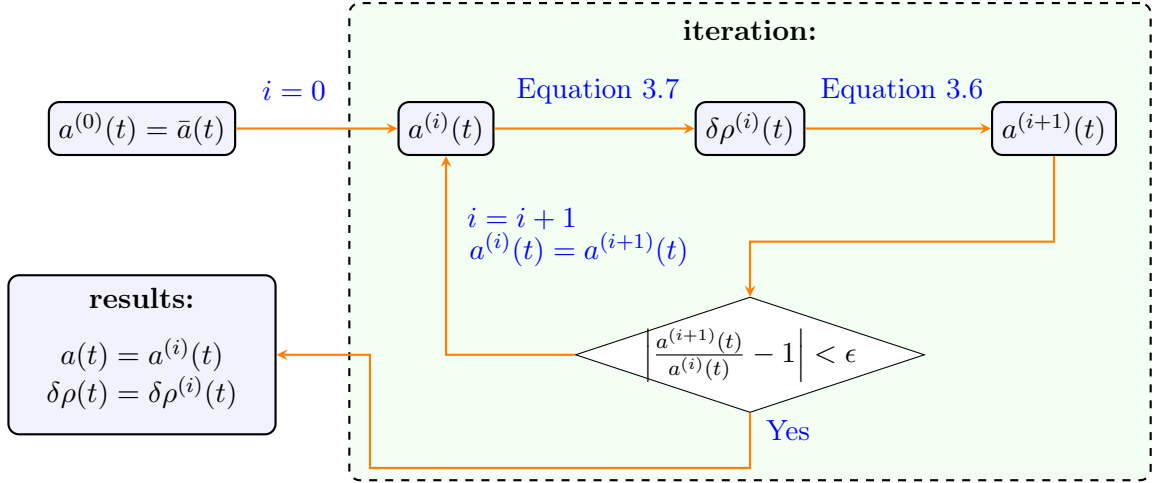

Moreover, 
taking the background dynamics of the contracting phase $\bar H=2/[3(1+w)t]$  (with $w$ being a constant) into \autoref{eq:linear},
one obtains 
\be
\label{eq:lin-con}
\ddot{\delta \rho}+\frac{2(8+3w)}{3(1+w)t} \dot{\delta \rho}+\frac{4(3+w)}{3(1+w)^2t^2}\delta \rho=0,~~~~~~~~~~~~~k \rightarrow 0 ~~\&~~ t\leq t_{\rm C}.
\ee 
The leading-order solution of \autoref{eq:lin-con} is $\delta \rho \propto  t^{-(3+w)/(1+w)}$. Hence, at the initial moment of the bounce phase ($t_{\rm C}$), one obtains 
\be
\label{eq:ini}
\dot {\delta \rho} \simeq -\frac{3(3+w)}{2} H_C \delta \rho. 
\ee
This will be used as an initial condition in the subsequent numerical computations of \autoref{eq:Raychaudhuri-2} during the bounce phase. 

\subsection{Numerical Realization}\label{Sec:numerical}

In this part, we perform numerical computations of \autoref{eq:Raychaudhuri-2} to obtain non-linear evolution of $\delta \rho$ during the bounce phase.

First of all, the results depends on the initial values of $\delta \rho$, $\dot {\delta \rho}$ and $k$.  
The initial $\delta \rho$ can be expressed as $3H_{\rm C}^2 \delta_i$, 
and the initial  $\dot {\delta \rho}$ can be fixed by \autoref{eq:ini}. 
Hence, $\delta_i$ and $k$ will be two input parameters in the following computations. 

Furthermore, combing \autoref{eq:a} and \autoref{eq:Raychaudhuri-2}, we have to handle a set of differential-integral equations, 
which are very complicated. 
Hence, we choose the method of iteration as a practical approach, as shown in \autoref{fig:flow}.  
It is seen that the initial $a(t)$ of the iteration is taken as the background one $\bar a(t)$ given by \autoref{eq:bounce-a}. 
The iteration stops at the $(i+1)$-th ($i \geq 0$) step, if the resulting $a^{(i)}(t)$ and  $a^{(i+1)}(t)$ nearly overlap.

\begin{figure}
  \centering
  \begin{overpic}[width=0.7\textwidth]{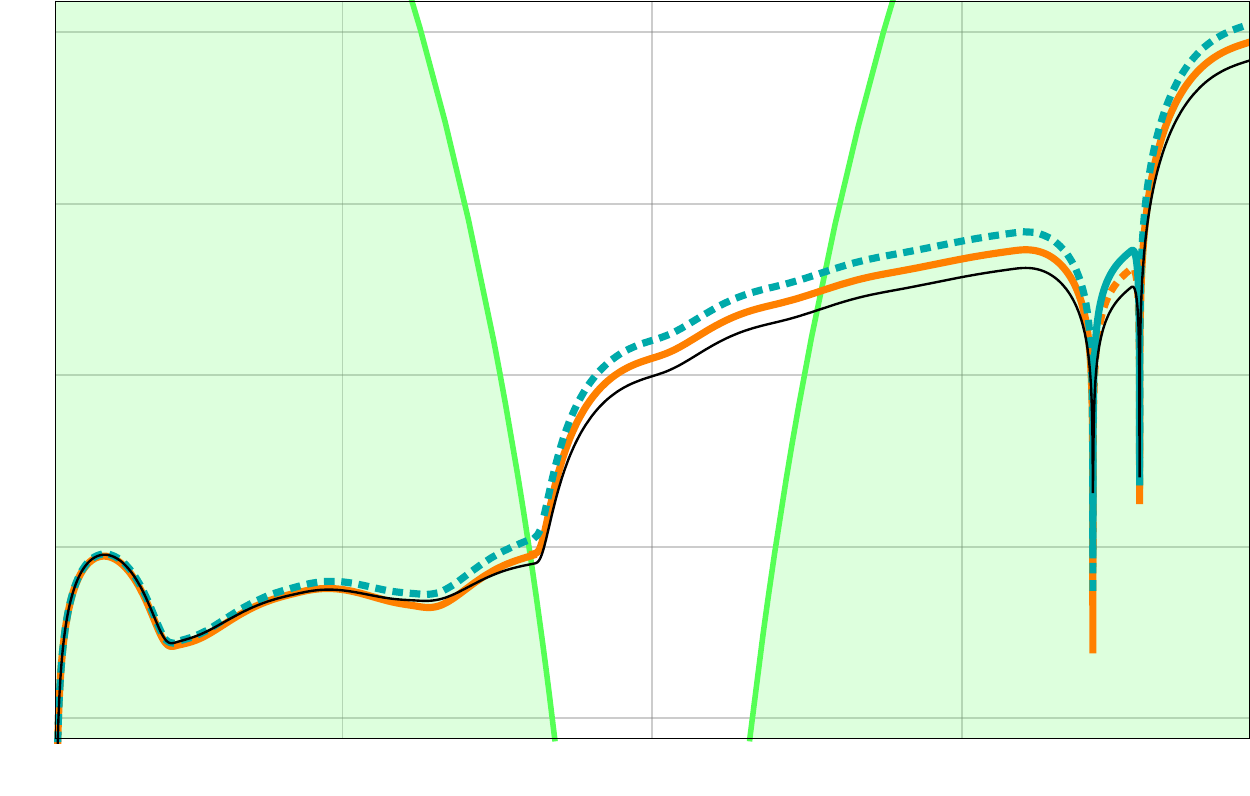}
  \put(48,-7){ $t / 10^{12}$}
  \put(-13,30){\begin{sideways} $|\delta \rho|$ \end{sideways}}
  \put(25,0){$-5$} \put(50,0){$0$}  \put(75,0){$5$} \put(0,0){$-10$} \put(98,0){$10$}
  \put(-5,5){$10^{-39}$} \put(-5,18){$10^{-37}$} \put(-5,32){$10^{-35}$} \put(-5,45){$10^{-33}$} \put(-5,58){$10^{-35}$}
  \end{overpic}\\
  ~\\
  ~\\
  \caption{An illustration for the evolution of $|\delta \rho|$ during the bounce phase with $H_{\rm BB}=-H_{\rm C}=6.23\times 10^{-12}$,
  $n_{\rm BB}=5$, $\gamma=11$, $c_s=1$. The orange and blue thick curves sketch the fluctuations with initial values $\delta_i=10^{-17}$ and $\delta_i=-10^{-17}$ respectively, and the wavenumber $k$ for both cases is taken as $0$.
  The dashed parts of the two lines denote the moments when the fluctuations  are under-dense $\delta \rho<0$.
  The black thin curve represents $|\delta \rho|$ given by the linear theory \eqref{eq:linear-m},
  and the green filled area represents the regime $|\delta \rho| \leq \bar \rho$.
  }\label{Fig:EvolveFig}
\end{figure}
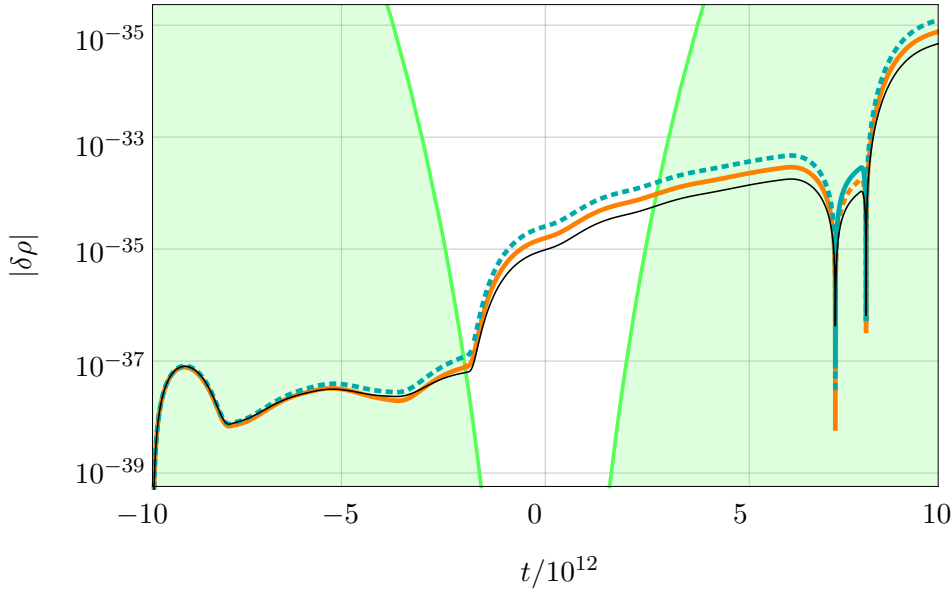

A specific result of the evolution for $\delta \rho$ during a bounce phase is shown in \autoref{Fig:EvolveFig}. 
Firstly, it is seen that the difference between linear and non-linear results becomes significant when $|\delta \rho|$ exceeds $\bar \rho$, as expected in \autoref{Sec:Intro}. Secondly, the evolution of $\delta \rho$ shows oscillating behaviors as mentioned in \autoref{Sec:linear} --- the initial cloud/void becomes a void/cloud during the span $t \in (7, 8)\times 10^{12}$, and then returns cloud/void later.
Furthermore, at the late stages $(t>3 \times 10^{12})$, 
$|\delta \rho|$ becomes much smaller than $\bar \rho$,
and the evolutionary behaviors of $\delta \rho$ from both non-linear and linear theories are nearly the same, 
except overall amplifications. 
It indicates that the non-linear effects become negligible in the late stage of bounce phase and the subsequent Big Bang phase. 
In the shown case by \autoref{Fig:EvolveFig},
the non-linear effects amplify $|\delta \rho|$ by factors of $1.64$ (for the case $\delta_i>0$) and $2.65$ (for the case $\delta_i<0$), respectively, 
compared with the result given by the linear theory.  
However, it is important to note that the non-linear effects may also reduce $|\delta \rho|$ in some cases,
hence the PBH abundance is not always enhanced. 
This brings requirements on the parameters of bounce phase to enhance PBH abundance, 
which will be discussed in \autoref{Sec:PBHs}.

\section{PBH Enhancement in Bouncing Cosmology}\label{Sec:PBHs}

\subsection{Probability of PBH formation} 

It is well known that the physical scales of fluctuations re-enter Hubble lengths, during the Big Bang phase\footnote{In this article, we only consider the PBH formation in the radiation dominated stage $z>z_{\rm eq}$.}. 
If a fluctuated region is compact enough at the Hubble re-entry moment ($\frac{1}{\sqrt{3}}k=\bar a \bar H$), 
i.e. with $\delta$ being larger than a threshold $\delta_c$, it will finally collapse to a PBH. 
The probability of PBH formation depends on the PDF of $\delta$. 
For a Gaussian PDF $\delta \in N(0, \sigma_{\rm re}^2)$, 
the probability is the well-known $\beta = \frac{1}{2}{\rm Erfc}(\delta_c/\sqrt{2}\sigma_{\rm re})$ \citep{Carr:1975, Carr:2010, Carr:2020} given by the Press-Schechter theory \citep{Press-Schechter:1974}. Here the Erfc denotes the complementary error function, and $\sigma_{\rm re}$ is the standard error of $\delta$ at the Hubble re-entry moment. 
However, in the bouncing scenario, 
the non-linear evolution of $\delta$ may be significant and the PDF may have strong NGs after the bounce point, as shown in \autoref{Fig:EvolveFig}. 
Hence the probability of PBH formation should be beyond the Press-Schechter formalism, and will be evaluated in the following.

As mentioned in \autoref{Sec:numerical},
the evolution of $\delta$ is linear during the Big Bang phase, 
which can be evaluated as $\delta \propto k^2/(\bar a^2 \bar H^2)$ before Hubble re-entry \citep{Cai:2018}. 
Therefore, the criterion $\delta \geq \delta_c$ at the Hubble re-entry is equivalent to 
\be
\label{eq:threshold}
\delta_{\rm BB} \geq \delta_{c, \rm BB} \equiv \frac{k^2}{3\bar a_{\rm BB}^2\bar H_{\rm BB}^2}\delta_c,
\ee
where $\delta_{c, \rm BB}$ denotes the threshold at $t_{\rm BB}$, and the value of $\delta_c$ is set as $0.37$ here \citep{Sasaki:2018}.  
Furthermore, the standard error of $\delta$ can be evaluated as $\sigma \simeq 4 k^2 P_\zeta^{1/2}/(27\bar a^2 \bar H^2)$ before Hubble re-entry \citep{Cai:2018}, where $P_{\zeta}$ is the power spectrum of the comoving curvature perturbation. 
Therefore, the standard error of density fluctuation at $t_{\rm BB}$ can be estimated as 
\be
\label{eq:standard}
\sigma_{\rm BB}=\frac{4 k^2}{27\bar a_{\rm BB}^2 \bar H_{\rm BB}^2} P_\zeta^{1/2}=\frac{4 k^2}{27\bar a_{\rm BB}^2 \bar H_{\rm BB}^2} A_s^{\frac{1}{2}} \left( \frac{k}{k_{\rm P}} \right)^{\frac{n_s-1}{2}},
\ee 
with the parameters $k_{\rm P}=0.05{\rm~Mpc}^{-1}$,  $A_s=2.1\times 10^{-9}$ and $n_s=0.965$ given by the CMB survey \citep{Planck:2018vyg}.

Moreover, since the NGs arise around the bounce point, 
it is reasonable to assume that the PDF of the initial density contrast is Gaussian $\delta_i \in N(0, \sigma_i^2)$. 
Here $\sigma_i$ is the standard error of $\delta_i$, which finally evolves to $\sigma_{\rm BB}$ through the bounce phase. 
Therefore, 
it is convenient to express the probability of PBH formation with respect to $\delta_i$, that
\be
\label{eq:beta}
\beta(k)={\mathop \int \limits_{\rm \autoref{eq:threshold}} }\frac{1}{\sqrt{2\pi} \sigma_i(k)} \exp\left[-\frac{\delta_i^2(k)}{2\sigma_i^2(k)}\right] d\delta_i(k).
\ee
Note that \autoref{eq:beta} has a similar form to the result in the Press-Schechter formalism \citep{Carr:2020}, 
except a modified criterion which is expressed in terms of $\delta_{\rm BB}$.

Given above, the probability of PBH formation in the bounce scenario can be calculated.

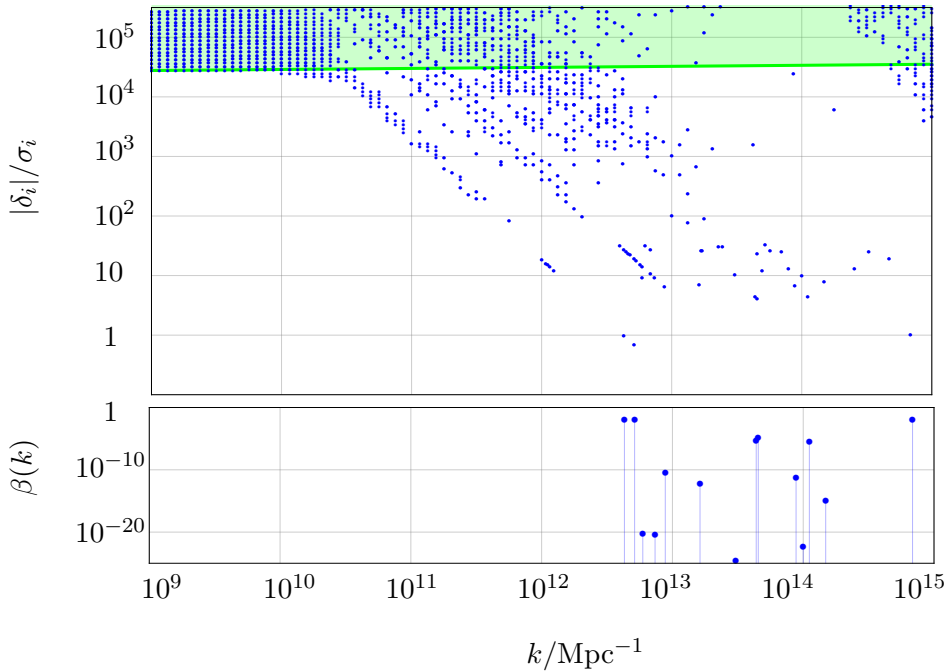
\begin{figure}
  \centering
  \begin{overpic}[width=0.8\textwidth]{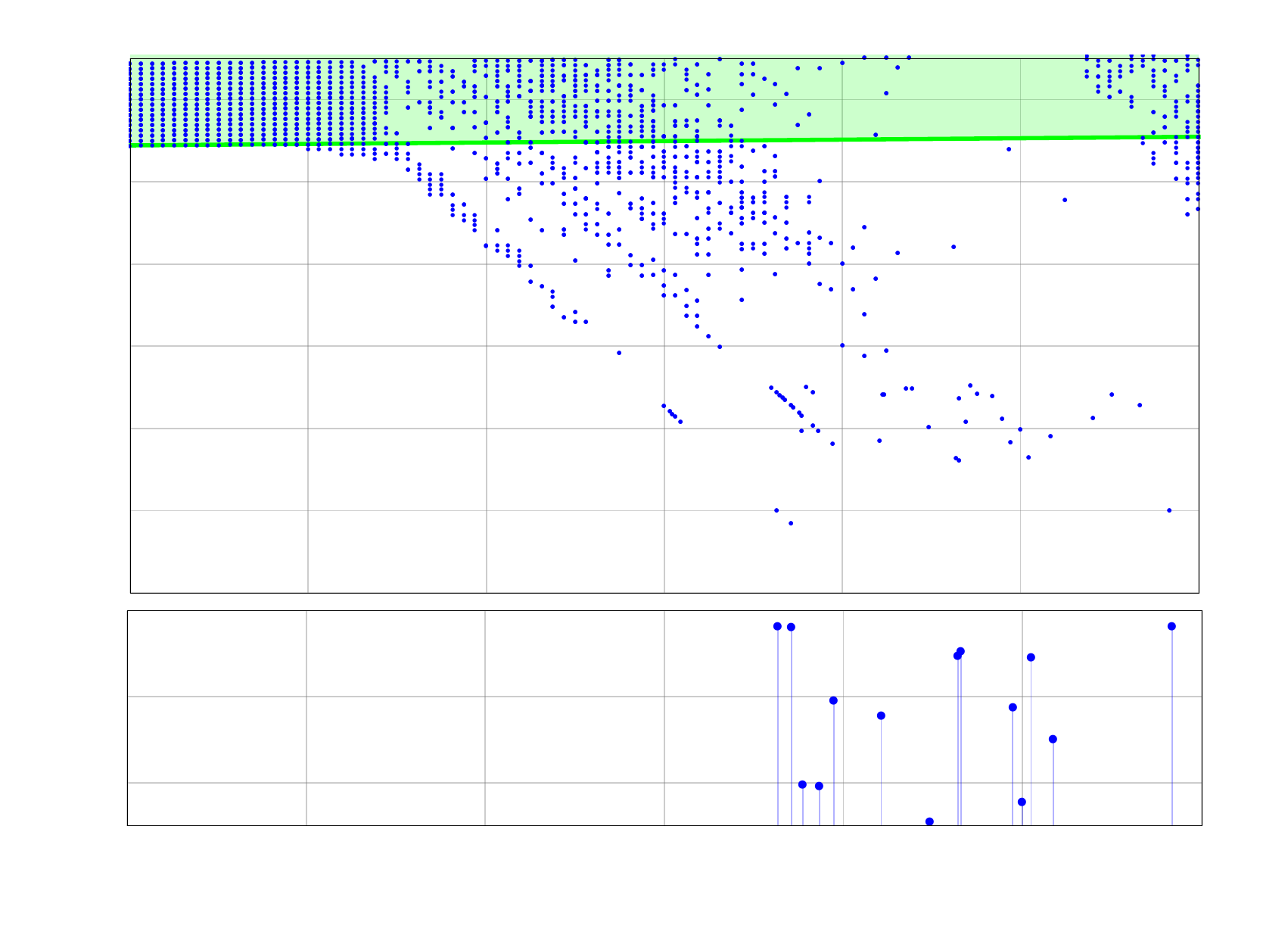}
  \put(50,-3){$k / {\rm Mpc}^{-1}$}
  \put(9,4){$10^9$} \put(22,4){$10^{10}$} \put(35,4){$10^{11}$} \put(49,4){$10^{12}$} \put(62,4){$10^{13}$} \put(75,4){$10^{14}$} \put(89,4){$10^{15}$}
  \put(-5,15){\begin{sideways} $\beta (k)$ \end{sideways}}
  \put(2,10){$10^{-20}$} \put(2,17){$10^{-10}$} \put(5,23){$1$}
  \put(-5,45){\begin{sideways} $|\delta_i|/\sigma_i$ \end{sideways}}
  \put(5,31){$1$} \put(4,38){$10$} \put(4,44){$10^2$} \put(4,51){$10^3$} \put(4,57){$10^4$} \put(4,63){$10^5$}
  \end{overpic}
  ~\\
  ~\\
  \caption{The upper panel illustrates the $(\delta_i, k)$ parameter space which leads to PBH formation and the lower panel shows the corresponding $\beta (k)$, 
  for a specific case with $H_{\rm BB}=-H_{\rm C}=5.68 \times 10^{-19}$, $\gamma=11$, $n_{\rm BB}=1$, $c_s=1$, 
  and $w=0$ before $t_{\rm C}$.  
  In the upper panel, 
  the green filled area stands for the result in linear theory (\autoref{eq:linear-m}), 
  and the blue dots represent the result in non-linear theory (\autoref{eq:Raychaudhuri-2}). 
  The numerical result is achieved under the precision $\Delta [\log_{10}(|\delta_i|/\sigma_i)]=2^{-6}$ and $\Delta (\log_{10} k)=2^{-7}$.}\label{Fig:deltaFig}
\end{figure}


\autoref{Fig:deltaFig} illustrates the  $(k, \delta_i)$ parameter space which can form PBHs, as well as the probability $\beta (k)$.
Firstly, it is clear that the non-linear effects during the bounce phase can significantly improve the probability of PBH formation. 
In details, the linear theory requires $|\delta_i|/\sigma_i \gtrsim 10^{4}$ for each $k$ to form PBHs, 
corresponding to negligibly small probabilities $\beta (k) \lesssim 10^{-2\times 10^7}$.  
Meanwhile, the non-linear theory enlarges the probabilities to $\beta(k) > 10^{-20}$, for some wavelengths.  
Secondly,  it is clear that the non-linear effects are significant only at small scales ($k \geq 10^{10}$ Mpc$^{-1}$ for the illustrated case), which will not affect the power spectrum and PDF of $\delta$ at CMB scales.  
Moreover, the profile of the parameter space forming PBHs is not continuous, consisting of discrete points, in the non-linear theory. 
This is because the non-linear evolutionary behavior of $\delta \rho$ is complicated. 
For example, an initially less denser region may finally evolve to a PBH, while an initial denser one may not.

\subsection{PBH Abundance}

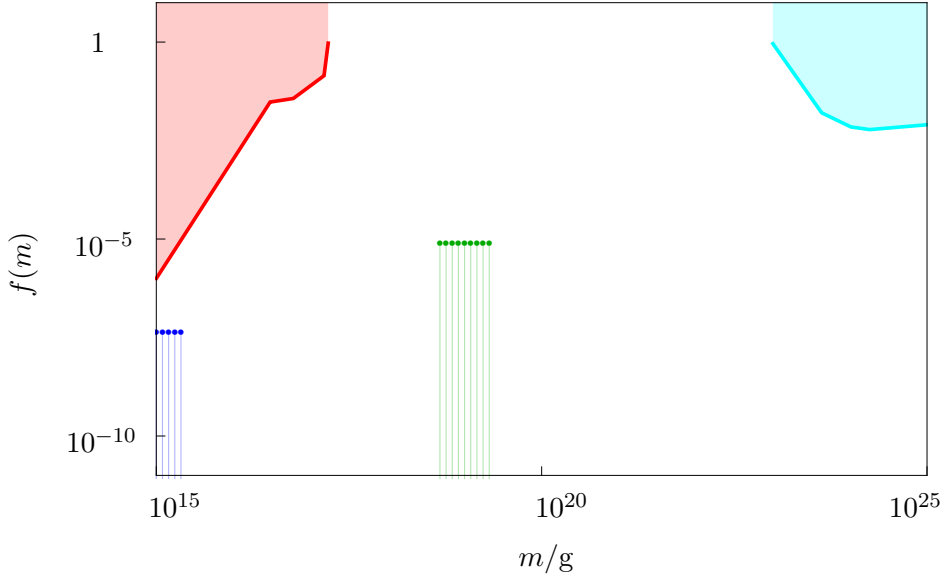
\begin{figure}
  \centering
  \begin{overpic}[width=0.7\textwidth]{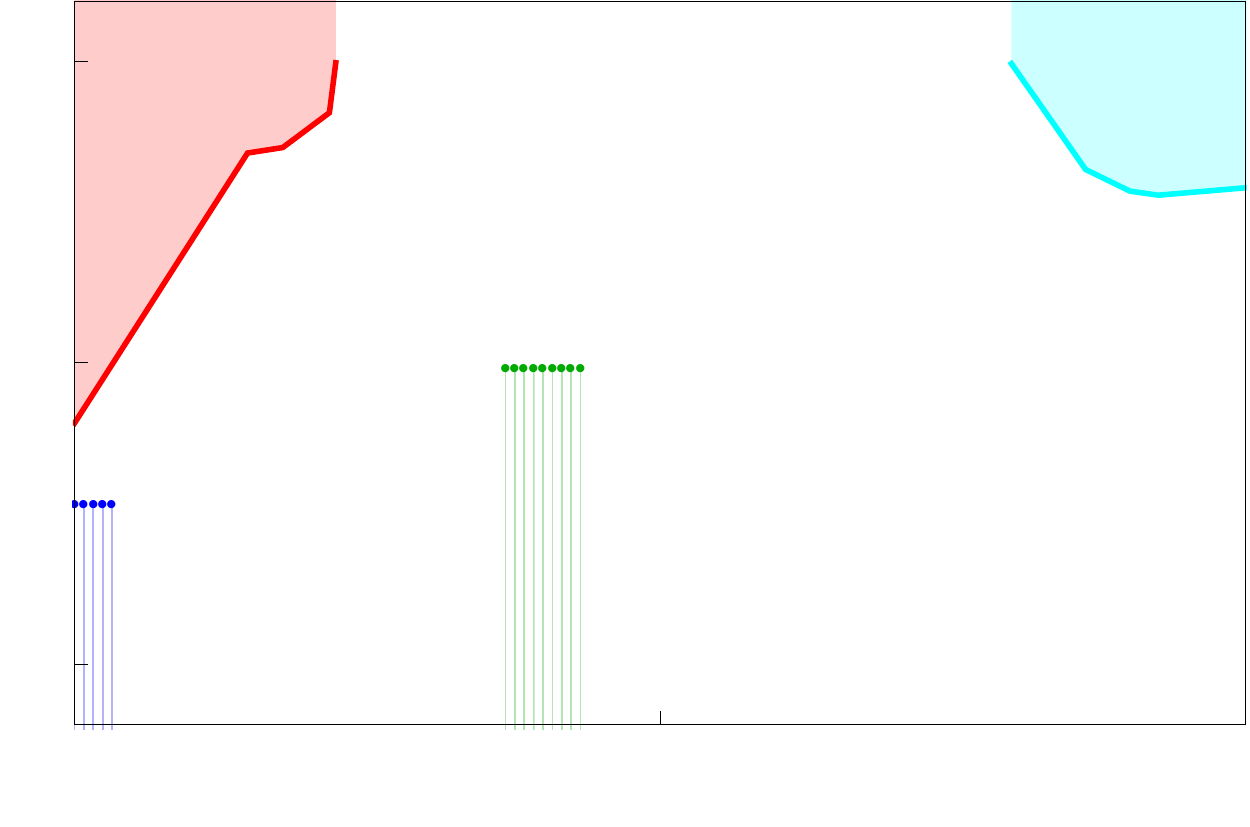}
  \put(50,-3){$m / {\rm g}$}
  \put(5,3){$10^{15}$} \put(52,3){$10^{20}$} \put(95,3){$10^{25}$}
  \put(-12,30){\begin{sideways} $f(m)$ \end{sideways}}
  \put(-5,11){$10^{-10}$} \put(-4,35){$10^{-5}$} \put(-2,60){$1$}
  \end{overpic}
  ~\\
  ~\\
  \caption{The PBH abundances $f(m)$ in bouncing cosmology. 
  The green dots represent the result given by the bounce phase with $\gamma=5$ and $n_{\rm BB}=0.31$, and the blue dots given by  $\gamma=5$ and $n_{\rm BB}=0.59$ --- the other parameters are the same as those in \autoref{Fig:deltaFig}. 
  The results are achieved under the numerical precision $\Delta [\log_{10}(|\delta_i|/\sigma_i)]=0.1$ and $\Delta (\log_{10} k)=0.04$.
  The red curve denotes the upper limit of $f(m)$ by observations of $\gamma$-ray background, Voyager positron flux and annihilation line radiation; the cyan curve denotes the upper limit of $f(m)$ by microlensing observations \cite{Carr:2020}.}\label{Fig:Fraction}
\end{figure}

Since $\beta (k)$ has been obtained, 
the PBH abundance and mass function will be evaluated in the following. 

First of all, the mass of PBH originating from a region with wavenumber $k$ is usually estimated as the mass inside the Hubble horizon at the re-entry moment \citep{Carr:2010, Carr:2020}
\be
\label{eq:mass}
m(k) \simeq \bar H^{-1}\big|_{k=\sqrt{3}\bar a \bar H} \simeq  10^{15} {\rm g} \left(\frac{k}{k_{\rm 15}} \right)^{-2},
\ee
with $k_{\rm 15}\simeq 10^{15}~{\rm Mpc}^{-1}$. 
Furthermore, the comoving number density of PBHs formed by fluctuated regions with wavenumber $k$ is $n(k) \simeq k^3 \beta(k)$ \citep{Carr:2020}. 
As a result, the total number of PBHs inside the observable Universe  (with comoving wavenumber $k_0 \simeq H_0$) is 
\be
\label{PBH-number}
N={\mathop \sum \limits_{k=i k_0}} \frac{n(k)}{k_0^3} \simeq \int \frac{k^3 \beta(k) dk}{k_0^4} 
\ee
with $i=1, 2, 3...$. 
Here the technique of box normalization ($k=i k_0$) is used. 

Moreover, since the heavy PBHs ($m>10^{15}$ g) contribute to part of dark matter components, 
the density fraction of PBHs against the total dark matter ($f\equiv \Omega_{\rm PBH}/\Omega_{\rm dm}$) is typically used to represent the PBH abundance \citep{IvanovPRD94, Carr:2016drx},
which is expressed as  
\be
\label{eq:fraction-tot}
f \simeq \int_{m(k')>{\rm 10^{15}g}} \frac{m(k')\, n(k')}{3H_0^2\Omega_{\rm dm}}\frac{dk'}{ k_0} \simeq 1.55\times 10^{36} \int_{k'<k_{\rm 15}} \frac{k'\beta(k') dk'}{k_{15}^2},
\ee
with $\Omega_{\rm dm} \simeq 0.264$ being the density fraction for the dark matter \citep{Planck:2018vyg}. 
Furthermore, the PBH abundance around a specific mass (actually in the PBH mass range $(m, m+\Delta m)$, with $\Delta m \simeq m$, or equivalently $\Delta k \simeq -k/2$ according to \autoref{eq:mass}) is also used \citep{Carr:2020},  which is 
\be
\label{eq:fraction}
f(m) \simeq 1.55\times 10^{36} \int \frac{k'\beta(k') dk'}{k_{15}^2} W(k'; k), 
\ee
with the window function 
\be
W(k'; k)= 
\bc
1,  &  k' \in (\frac{k}{2}, k) ~\&~k'<k_{\rm 15}\\
0,  &  {\rm~else}
\ec .
\ee

\autoref{Fig:Fraction} illustrates the PBH abundances $f(m)$ given by specific bounce phases.
It is seen that 
the bounce phase, at lest for the specific cases, can sufficiently improve the PBH abundance and do not break the observational constraints. 
Hence it can serve as an potential mechanism of PBH formation.

\subsection{Constraints of Bounce Phase through PBH abundance}\label{Sec:Constraints}

Since the PBH abundance depends on the background dynamics of bouncing scenario, as shown in \autoref{Fig:Fraction},
the PBHs can conversely provide a probe to the parameters of bounce phase, which includes 
$H_{\rm C}$, $H_{\rm BB}$, $n_{\rm BB}$, $\gamma$ and $c_s$. 
Note that, the parameters $H_{\rm C}$, $H_{\rm BB}$ and $c_s$ can be effectively constrained combining CMB observations \citep{Cai:2012va}, 
but $n_{\rm BB}$ and $\gamma$ are generally poorly measured. 
Therefore, the constraints of $n_{\rm BB}$ and $\gamma$ by PBHs are worthwhile. 

\begin{figure}
  \centering
  \begin{overpic}[width=0.7\textwidth]{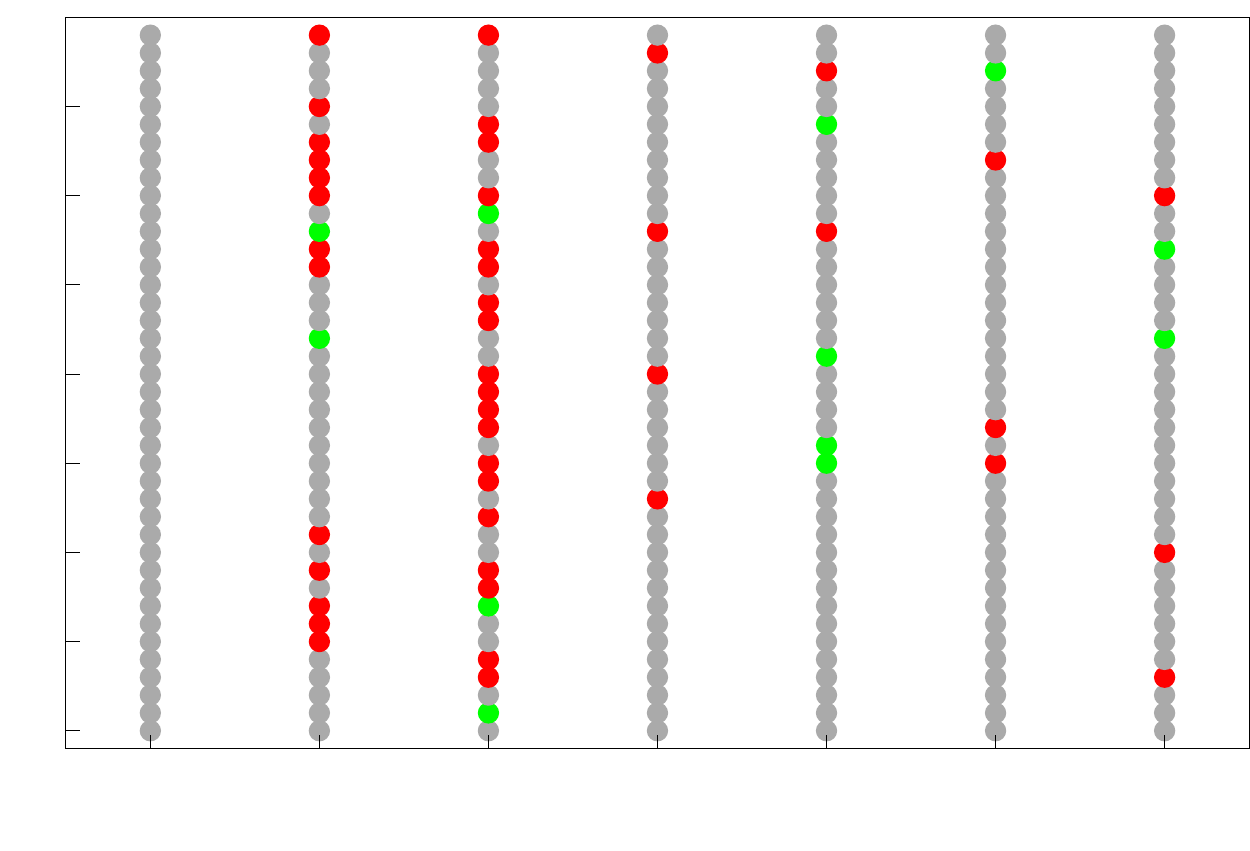}
  \put(50, -2){$\gamma$} 
  \put(11,3){$1$} \put(24,3){$3$} \put(38,3){$5$} \put(51,3){$7$} \put(65,3){$9$} \put(77,3){$11$} \put(90,3){$13$}
  \put(-10, 35){\begin{sideways}$n_{\rm BB}$ \end{sideways}}
  \put(-2,8){$0.3$} \put(-2,23){$0.4$} \put(-2,38){$0.5$} \put(-2,51){$0.6$} \put(-2,65){$0.7$}
  \end{overpic}
  ~\\
  ~\\
  \caption{Constraints of bounce phase parameters $n_{\rm BB}$ and $\gamma$, with $H_{\rm BB}=-H_{\rm C}=6.23 \times 10^{-12}$ and $c_s=1$. 
  The red dots represent the excluded bounce phases, which generate too many PBHs.
  The gray dots denote the disfavored bounce phases, which cannot generate PBHs effectively ($N<1$). 
  The green dots stand for the favored bounce phases. 
  The results are achieved under the numerical precision $\Delta [\log_{10}(|\delta_i|/\sigma_i)]=0.1$ and $\Delta (\log_{10} k)=0.04$.}\label{Fig:Constrain}
\end{figure}
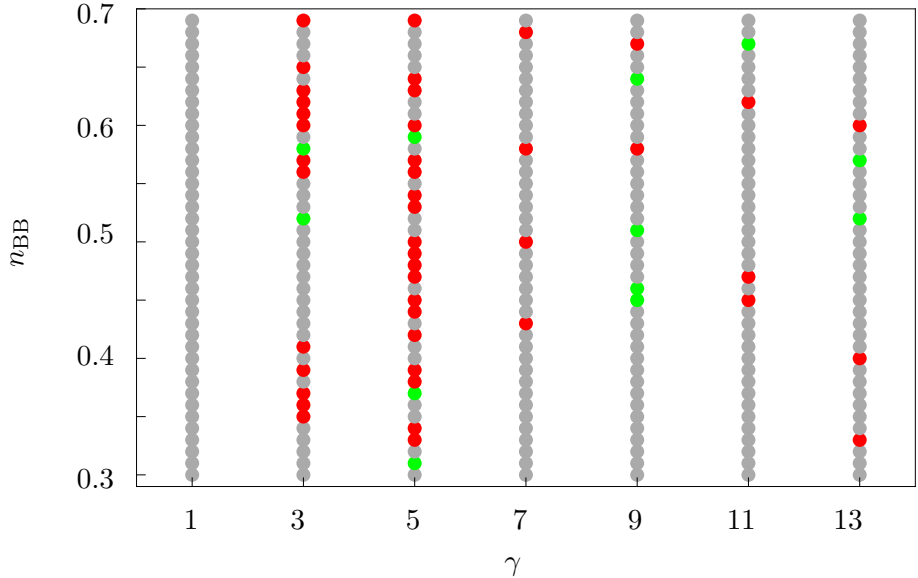

The PBH abundance $f(m)$ in a wide mass range $(10^{15}$-$10^{55})$ g has been constrained \citep{Carr:2020}. 
For a bounce model with parameters $n_{\rm BB}$ and $\gamma$, 
it will be ruled out if too many PBHs are generated, i.e. 
in the following two cases: 
(1) the abundance $f(m)$ calculated by \autoref{eq:fraction} breaks an observational constraint; 
(2) the total abundance $f$ given by \autoref{eq:fraction-tot} is larger than $1$. 
Furthermore, if the PBH number $N$ given by \autoref{PBH-number} is smaller than $1$, 
the corresponding bounce phase cannot generate PBHs effectively. 
In this case, although the bounce phase is safe in the observational constraint, 
it is disfavored in the motivation of PBH formations. 
Given above, only the bounce phases neither generating too many PBHs nor too few PBHs are favored. 

The constraints of $n_{\rm BB}$ and $\gamma$ are illustrated in  \autoref{Fig:Constrain}. 
Firstly, it is clear that the bounce phase can be constrained by PBHs. 
Furthermore, the favored bounce phases (green dots) distribute discretely in the parameter space, 
and the rule of the distribution is still unknown.
This is probably because the numerical precision in our calculations is not high enough. 
Note that an increasing precision may improve the results significantly.

\section{Conclusions and Outlook}\label{Sec:Conclusion}

In this article, we apply the non-linear evolution of density fluctuation around bounce point to enhance PBH abundance.  
Note that the non-linear effects naturally exist in the relativistic bouncing scenario,
hence the PBH formation does not require an extra physical mechanism. 
Our results in \autoref{Sec:PBHs} indicate that it is plausible to produce PBHs sufficiently through bounce phase,
which improve the pessimistic conclusions in the earlier researches \cite{Quintin_2016, CHEN2017561}. 
We express the PBH abundance through the parameters of bounce phase, 
and  the PBH abundance can also provide a probe to the parameters $\gamma$ and $n_{BB}$, 
which are usually difficult to be measured by the CMB or large scale structure (LSS) observations. 
Therefore, the PBH observations may provide a complementary to the future surveys of CMB and LSS.

The current work can be extended in the following three aspects.

Firstly, the non-linear \autoref{eq:Raychaudhuri-2} is so complicated that the numerical errors may have considerable impacts on the results.
For example, the linear \autoref{eq:linear-m} and the non-linear \autoref{eq:Raychaudhuri-2} actually lead to different transfer functions at $\delta_i \to 0$ due to the numerical errors, with the deviation being several orders of magnitude. 
To reduce the impacts from numerical errors, the linear transfer function in this work is actually achieved by solving the non-linear \autoref{eq:Raychaudhuri-2} at $\delta_i=10^{-40}$ \footnote{As we have found, for $|\delta_i|\sim 10^{-40}$, the linear and non-linear equations approximately yield the same results; if $|\delta_i|<10^{-40}$, the results will be dominated by numerical errors. }. 
This algorithm should be further tested or improved in the future. 
Moreover, as mentioned in \autoref{Sec:Constraints}, the results will change significantly as the numerical precision increases, so will be the time of the computations. 
Therefore, our results can only serve as a rough estimation, and it is worthwhile to spend more time improving the precision in the follow-up researches.

Secondly, the subsequent evolution of the formed PBHs, including evaporation \citep{Page&Hawking:1976}, accretion and merger \citep{Chen_2018},
are not considered in this article, which will change the abundance and mass function for the PBHs of interest. 
Hence we also plan to include these effects and update the constraints of bounce phase in the future.

Moreover, since the PBH enhancement in this work has a close connection to the NGs of density fluctuation, 
the PBH abundance can be expressed in terms of the parameters of NGs, such as $f_{\rm NL}$ and $g_{\rm NL}$ etc. \citep{Young_2013, Passaglia2019, Luca_2019}. 
This is another point worth being investigated in the follow-up researches. 
Additionally, 
the NG parameters $f_{\rm NL}$ and $g_{\rm NL}$ have been constrained by CMB surveys \citep{Planck:2019kim}.
Hence the bounce phase can be constrained though the joint of the PBH abundance and observational results of $f_{\rm NL}$ and $g_{\rm NL}$.


\acknowledgments
We are grateful to Bernard Carr, Robert Brandenberger, Jerome Quintin, Linhua Jiang, Huiyuan Wang, Chao Chen, Qianhang Ding, Yi Wang, Yan Wang, Yiqiu Ma and Daiqin Su for valuable guidance, discussions and comments.  J. W. C. acknowledges the support from China Postdoctoral Science Foundation under Grant No. 2021M691146. SFY is supported by the Disposizione del Presidente INFN n.24433 ``{\it Quantum Fields for Gravity, Cosmology and Black Holes}'' in INFN Sezione di Milano.

\bibliography{main}{}
\bibliographystyle{JHEP}

\end{document}